\documentclass{article}

\usepackage{arxiv}

\usepackage[utf8]{inputenc} 
\usepackage[T1]{fontenc}    
\usepackage{hyperref}       
\usepackage{url}            
\usepackage{booktabs}       
\usepackage{amsfonts}       
\usepackage{nicefrac}       
\usepackage{microtype}      
\usepackage{lipsum}

\usepackage{makecell}
\usepackage{multirow}
\usepackage{amsmath}
\usepackage{amsthm}
\usepackage{graphicx}
\usepackage{subcaption}
\usepackage{cite}
\usepackage{authblk}
\usepackage{float}

\newtheorem{lemma}{Lemma}
\newtheorem{protocol}{Protocol}

\providecommand{\keywords}[1]{\textbf{\textit{Keywords:}} #1}

\DeclareMathOperator{\VaR}{VaR}

\graphicspath{{figures/}} 
\title{Dynamic cyber risk estimation with Competitive Quantile Autoregression}

\author[1]{Raisa Dzhamtyrova}
\author[1, 2] {Carsten Maple }

\affil[1]{\footnotesize The Alan Turing Institute}
\affil[2]{\footnotesize Secure Cyber Systems Research Group, WMG, University of Warwick}
\affil[ ] {\textit \url{rdzhamtyrova@turing.ac.uk} \hspace*{0.3cm} \url{cm@warwick.ac.uk}}

\begin{document}
\maketitle

\begin{abstract}
The increasing value of data held in enterprises makes it an attractive target to attackers. The increasing likelihood and impact of a cyber attack have highlighted the importance of effective cyber risk estimation.  We propose two methods for modelling Value-at-Risk (VaR) which can be used for any time-series data. The first approach is based on Quantile Autoregression (QAR), which can estimate VaR for different quantiles, i.~e.~confidence levels.  The second method, we term Competitive Quantile Autoregression (CQAR), dynamically re-estimates cyber risk as soon as new data becomes available. This method provides a theoretical guarantee that it asymptotically performs as well as any QAR at any time point in the future. We show that these methods can predict the size and inter-arrival time of cyber hacking breaches by running coverage tests. The proposed approaches allow to model a separate stochastic process for each significance level and therefore provide more flexibility compared to previously proposed techniques. We provide a fully reproducible code used for conducting the experiments.
\end{abstract}


\section{Introduction}
The prevalence and impact of cyber attacks on organisations are increasing at an alarming rate. Risk estimation is an important task for any company or institution, allowing them to predict and assess adverse events which can lead to financial and reputation losses, enabling them to plan for and mitigate against these threats through effective risk management. 

Kaplan defines risk to be a set of triplets, which consist of a risk scenario description, the probability of that scenario, and the consequence or evaluation measure of that scenario, i.e., a measure of damage \cite{kaplan1981riskdef}. Another definition of risk is provided by Holton, in which the risk comprises two components: uncertainty and exposure \cite{holton2004riskdef}. Indeed all definitions of risk require some form of assessment of the likelihood of adverse events and their severity. A recent report by the Department of Homeland Security \cite{jones2018riskmetrics} provides a survey of risk metric frameworks and risk models. One of the quantitative risk metrics described in the report is Cyber Value-at-Risk (VaR), an adaptation of the financial VaR, for the quantification of cyber security risk. VaR is one of the most important risk measurements in finance and involves measuring the maximum loss over a preset horizon with a pre-defined confidence level \cite{hull6th}. VaR has now found various applications in cyber security areas. For example, Factor Analysis of Information Risk (FAIR), considered ``an international standard information risk management model'', is based on VaR. FAIR is defined as ``a standard Value-at-Risk model for information and operational risk that helps information risk, cyber security and business executives measure, manage, and communicate on information risk in a language that the business understands, dollars and cents'' \cite{jones2018riskmetrics}. In \cite{peng2016attacks} VaR is used to estimate the probability of extreme cyber attacks over a pre-defined period of time. The paper \cite{raugas2013cybervar} proposes a model to quantify the monetary VaR due to cyber threats based on the Bayesian networks. The detailed model describes the example attack graph of unauthorized access to intellectual property. In this paper, we propose a new methodology of estimation of VaR for cyber events.

In this paper, we aim to provide a framework that can model risks \emph{dynamically} and re-estimate cyber risk when new data becomes available. Many current risk methods are based on manual risk analysis during the system's design process. Some of the examples of traditional qualitative methods include scenario analysis and questionnaires, which are heavily dependent on experts' subjective opinions. On the other hand, quantitative risk methods are usually based on unreliable data, and therefore their precision is prone to errors \cite{taubenberger2011problems}. As a result, there is a lack of current research on dynamic cyber risk estimation. Of the work that has been proposed for dynamic risk modelling, a number of approaches are based on Hidden Markov Models (HMM). The paper \cite{arnes2005risk} proposes a real-time risk estimation method, which aggregates data from several intrusion detection systems allowing dynamic estimation of systemic risk using HMM. The paper \cite{li2018dynamicrisk} developed a method to dynamically model the risks of users' activity patterns in social networks. The approach is based on HMM and Bayesian Risk Graph model. Unlike the previous approaches, we do not model the dynamics of the system states with HMM. Instead, we focus on time-series data and propose a new method for dynamic estimation of VaR. System monitoring is essential to effective risk governance. The monitored data is usually a different kind of time-series, such as various sensor data, login data, and intrusion and hacking attempts. From a risk perspective, it is critical to estimate the probability of extreme events. For example, we do not want to predict the mean or the median of hacking attempts over a pre-defined period. Instead, we aim to assess the maximum number of hacking attempts with the desired confidence. For this purpose, we suggest to model VaR as a quantile of time-series, where each quantile corresponds to the desired confidence level. These values of VaR can also be translated into the monetary equivalent. For example, if we assume that each cyber hacking attempt costs one pound for a company, we can estimate the budget allocation which should be devoted to security defence. The proposed method builds upon the Weak Aggregating Algorithm for Quantile Regression (WAAQR), proposed in \cite{dzhamtyrova2020quantile} and is adapted to the case of time-series forecasting. 

We apply the proposed approach to the prediction of cyber hacking breaches. The data is taken from the report of the Privacy Rights Clearinghouse (PRC), which contains the chronology of the reported data breaches since January 2005~\footnote{\label{clearinghouse} \url{https://privacyrights.org/data-breaches}}. 
This benchmark dataset has been used by a number of other researchers in establishing the efficacy of their work \cite{edwards2016breaches, xu2018breaches }.  
The analysis of data breaches is attracting research activity lately, given the importance of the topic. Some of this work suggests that data breaches can be modelled using a variety of distributions. In \cite{hubbard2016risk} the authors suggest using the beta distribution for estimating the probability of data breaches based on industry data. After estimating the probability of data breaches, the VAR is modelled with the Monte-Carlo simulation, which gives a forecast of the possible losses. The paper \cite{edwards2016breaches} investigates the PRC data set from the period between January 2005 and September 2015. The study examined over 20 different distributions, such as log-normal, power-law, generalised Pareto to determine which provided the best fit for the size of the data breach. To model the breach frequencies, the authors investigated a number of discrete distributions, such as Poisson, binomial, and negative binomial. The results suggest that neither the size nor the frequency of data breaches has increased over the period under consideration. Furthermore, the study proposes to model the daily frequency of breaches using the negative binomial distribution, whereas breach sizes are best described by the log-normal family of distributions. It is, of course, possible that the nature of data breaches has changed significantly in the era of increasing data connectivity. The paper \cite{xu2018breaches} analyses the PRC data set with a focus on hacking breach incidents. Their analysis shows that both the inter-arrival time and the size of hacking breaches reveal significant auto-correlation and partial auto-correlation,  suggesting that the breaches can be modelled with stochastic processes. The paper estimates the inter-arrival times with the autoregressive conditional mean model, whereas the breach sizes are estimated with ARMA(1, 1)-GARCH(1, 1). The authors also show that there is a positive correlation between inter-arrival times and sizes of cyber incidents, and describe this dependence by a particular copula.

In this paper, we propose a new framework for dynamic estimation of VaR. Though the proposed methods can be used to predict any types of time-series, we perform our experiments on the PRC data report. The reasons are that this dataset contains one of the largest cyber events data available online, it is regularly updated, and it was studied before in the literature. Our analysis closely resembles the analysis in \cite{xu2018breaches}, however, we propose different modelling approaches of hacking breaches. First, our methods can be applied to any kind of time-series data. Second, the analysis in \cite{xu2018breaches} models the mean of inter-arrival times and sizes, and then VaR is found by simulating 10000 samples based on the estimated copula. Instead, we suggest that each quantile of inter-arrival times and sizes of cyber incidents can be modelled with separate stochastic processes. Though we do not investigate the relationship between inter-arrival times and sizes of breaches, we argue that the proposed methods are more flexible in comparison to previous research as they make fewer assumptions on the nature of the data, since each quantile of breach size or inter-arrival time can be modelled with a separate stochastic process. In our experiments, we first show that we can apply Quantile Autoregression (QAR) \cite{koenker2006qar} to estimate VaR of hacking breaches. The Basil Committee recommends assessing the quality of the VaR models by running some form of backtesting. Standard backtesting methods include the Kupiec unconditional coverage test \cite{Kupiec1995VaR} and the Christoffersen conditional coverage test \cite{Christoffersen1998}. We apply both tests to assess the performance of QAR. The results show that for breach size QAR fits well, and for an inter-arrival time, it rejects the null hypothesis of the conditional coverage test of violation occurrence for one considered quantile. We then propose a new framework, Competitive Quantile Autoregression (CQAR), which improves the prediction of hacking breach inter-arrival times. 

The proposed method CQAR is based on the competitive prediction approach, where one algorithm `competes' with other predictive algorithms. The goal is to provide a strategy that can guarantee a performance close to the best predictive models. To solve the problem of competitive prediction, the Aggregating Algorithm (AA) was proposed in \cite{vovk_aggr}. The AA mixes the predictions of a number of models in a similar manner to the Bayesian method, where the prediction is calculated based on the model's prior distribution and the data likelihood. Furthermore, the AA guarantees that the loss of the resulting mixing strategy is as small as the best model's plus a constant for any time point in the future. The Weak Aggregating Algorithm (WAA) \cite{Kalnishkan2008WAA} was proposed as an alternative for the AA, which provides better theoretical guarantees for some loss functions, such as the pinball loss, which we consider in this paper. In the general case, both the AA and the WAA mix and compete with a finite number of algorithms. 

It is possible to construct strategies that combine infinite classes of functions and provide theoretical guarantees compared to these classes. The Aggregating Algorithm for Regression chooses the competitor strategies to be all linear functions \cite{vovk_cols}. The resulting strategy asymptotically performs as well as any linear regression in terms of the cumulative square loss. A similar approach is undertaken in \cite{dzhamtyrova2020quantile}, where the authors propose the Weak Aggregating Algorithm for Quantile Regression \cite{dzhamtyrova2020quantile}. The strategy is a Bayesian mixture, which combines an infinite pool of quantile regressions, and asymptotically predicts as well as any of them in terms of the cumulative pinball loss. The algorithm was previously applied to probabilistic forecasting of renewable energy and showed a good performance on real data.
The proposed algorithm CQAR is built on the WAAQR algorithm and is adapted to time-series forecasting. Instead of mixing a class of quantile regressions, we suggest combining a class of QAR. It also has the property that it asymptotically predicts as well as any QAR. We provide the pseudo-code of CQAR, which uses Metropolis-Hastlings sampling  \cite{Andrieu2003MCMC} to calculate its predictions, however, it can be substituted with any other sampling algorithm. We show that CQAR produces better results in comparison to QAR for estimating VaR of hacking breach inter-arrival times. Another advantage of CQAR is that it re-estimates cyber risks dynamically after new observations become available. We also plot the average regret between CQAR and the best QAR depending on time and show that it goes to zero as time increases. This empirically confirms the theoretical guarantees of the method and shows that CQAR is asymptotically as good as the best QAR which was trained on the training data set.

\section{Contributions}
Our first contribution is a new analysis and adaptation of QAR for calculating cyber VaR. To the best of our knowledge, it was not done before. The method can be applied to any time-series data. It is common to predict the mean or median values of time-series. Some research also focuses on the prediction of extreme values. This analysis provides a new way to model extreme values that also comes with the desired confidence level. QAR allows to model VaR for each confidence level with a separate stochastic process, and hence allows more flexibility compared to previously proposed approaches in the literature.

The second contribution is a new dynamic risk estimation method, Competitive Quantile Autoregression (CQAR). There is a lack of research on dynamic cyber risk estimation. CQAR allows to re-estimate cyber risk at each time step when new data becomes available and works for any time-series data. An important property of this approach is its theoretical guarantee that it asymptotically predicts as well as the best QAR. The theoretical performance guarantees provide confidence in the prediction as they will hold for any new unseen data, while at the same time the method allows adapting to a changing environment. As with QAR, CQAR is also more flexible as it models each quantile with a separate stochastic process.

The third contribution is the modelling of cyber hacking breaches with the proposed methods. We show that both QAR and CQAR can be used to estimate VaR of cyber hacking breaches' sizes and inter-arrival times. The coverage tests show a good fit of both approaches. We show that CQAR provides better results for modelling hacking breaches' inter-arrival times compared to QAR. We also illustrate the behaviour of the average regret between CQAR and QAR during the time and show that it conforms to the theoretical bounds of CQAR. The fully reproducible open-source code of our implementation is available at GitHub~\footnote{\label{cyber_risk} \url{https://github.com/alan-turing-institute/dynamic_cyber_risk}}.

\section{Risk estimation with Quantile Autoregression}
VaR is a widely used risk measurement in finance. $\VaR_{\alpha}$ is defined as the loss corresponding to the $\alpha$-quantile of the distribution of the gain in the value of the portfolio over the next $N$ days (Chapter 21.1 in \cite{hull6th}). In finance, VaR provides an estimate of the maximum loss for a certain confidence level and is important for budget allocation and financial reserves. Analogously, in cyber security, we want to estimate possible losses of extreme cyber events, such as cyber attacks and subsequent data losses. Accurate forecasting of these adverse events can allow an adaptation of risk mitigation strategies and better financial planning. 

Let the outcomes have a cumulative distribution $F_Y(z)$, then we define
\begin{equation} \label{eq:VaR_definition}
	\VaR_{\alpha} =  \inf\{z: F_Y(z) \ge \alpha\}
\end{equation}
as the $\alpha$-quantile of $Y$. Then we can estimate $\VaR_\alpha$ as $\alpha$-quantile of outcomes. 

QAR, proposed in \cite{koenker2006qar}, allows to model each quantile of outcomes with a separate autoregressive process. Let time-series $y_t$ to be the $p$-order autoregressive process:
\begin{equation} \label{eq:ar_process}
	y_t = \theta_0(U_t) + \theta_1(U_t) y_{t-1} + \dots + \theta_p(U_t) y_{t-p},
\end{equation}
where $\{U_t\}$ is a sequence of i.i.d. standard uniform random variables. We want to estimate the coefficients $\theta_j$, which are unknown functions $[0, 1] \rightarrow \mathbb{R}$.
The $\alpha$th conditional quantile of $y_t$ is:
\begin{equation} \label{eq:qar_definition}
	Q_{y_t}(\alpha| y_{t-1}, y_{t-2}, \dots, y_{t-p}) = \theta_0(\alpha) + \theta_1(\alpha) y_{t-1} + \dots + \theta_p(\alpha) y_{t-p}.
\end{equation}
Equation (\ref{eq:qar_definition}) can be rewritten in analogous to the definition of quantile regression in \cite{Koenker1978QR}:
\begin{equation} \label{eq:qar_definition2}
	Q_{y_t}(\alpha| \mathcal{F}_{t-1}) = x_t^\prime \theta(\alpha),
\end{equation}
where $x_t = (1, y_{t-1}, \dots, y_{t-p})^\prime$, $\theta = (\theta_0, \theta_1, \dots, \theta_{t-p})^\prime$, and $\mathcal{F}_{t-1}$ is the $\sigma$-field generated by $\{y_s, s \le t\}$.

The coefficients $\theta(\alpha)$ in (\ref{eq:qar_definition2}) are found by minimising the following expression:
\begin{equation} \label{eq:min_pinball}
	\min_{\theta \in \mathbb{R}^{p+1}} \sum_{t} \lambda(y_t, x_t^\prime \theta),
\end{equation}
where $\lambda(y, \gamma)$ is the pinball loss function:
\begin{equation}\label{eq:lossPinball}
	\lambda(y, \gamma) = \begin{cases}
		\alpha (y - \gamma), & \text{if $y \ge \gamma$}  \\
		(1-\alpha) (\gamma - y), & \text{if $y < \gamma$} 
	\end{cases}.
\end{equation}
\section{Framework of competitive prediction}
In this section, we describe the framework of competitive prediction. In this framework, a {\em learner} plays a {\em game} $\mathfrak{G}$ against other prediction strategies and a {\em nature}, which reveals the true outcomes. 
A game $\mathfrak{G} = \langle \Omega, \Gamma, \lambda \rangle$ is a tuple with the space of outcomes $\Omega$, decision space $\Gamma$, and a loss function $\lambda$. In this paper, we consider $\Omega = \Gamma = \mathbb{R}$, and $\lambda$ to be the pinball loss, defined in (\ref{eq:lossPinball}) for $\alpha \in (0, 1)$.

The learner works according to the following protocol:
\begin{protocol}~
	\label{protocol1}
	\begin{tabbing} 
		\quad\=\quad\=\quad\=\quad\=\quad\kill
		for $t=1,2,\ldots$\\
		\>\> nature announces signal $x_{t} \subseteq \mathbb{R}^{p+1}$\\
		\>\> learner outputs prediction $\gamma_{t} \in \Gamma  $\\
		\>\> nature announces outcome $y_{t} \in \Omega  $\\
		\>\> learner suffers loss  $\lambda(y_{t}, \gamma_{t})$\\
		end for
	\end{tabbing}
\end{protocol}
Before seeing the true outcome $y_t \in \Omega$, the learner needs to make a prediction $\gamma_t \in \Gamma$, based on a signal $x_t$, which is announced by nature. After seeing the true outcome $y_t$, the learner's loss $\lambda(y_t, \gamma_t)$ can be calculated.

In this paper, we assume that the outcomes follow the $p$-order autoregressive process defined in (\ref{eq:ar_process}). The learner makes a prediction $\gamma_t$ based on the signal $x_t = (1, y_{t-1}, \dots, y_{t-p}) \in \mathbb{R}^{p+1}$. For ease of notation, we replace $\theta(\alpha)$ with $\theta$. Let us denote $\xi_t(\theta)$ to be the prediction (\ref{eq:qar_definition2}) of QAR(p):
\begin{equation} \label{eq:expert_prediction}
	\xi_t(\theta) = x_t^\prime \theta.
\end{equation}

We denote the cumulative loss of the learner at step $T$ as:
\begin{equation*}
	L_T :=  \sum_{t=1}^T \lambda (y_t, \gamma_t) = \sum_{\substack{t = 1, \dots, T:\\{y_t > \gamma_t}}} \alpha |y_t - \gamma_t| +
	\sum_{\substack{t = 1, \dots, T:\\ y_t < \gamma_t}} (1-\alpha) |y_t - \gamma_t|.
\end{equation*}


The cumulative loss of the prediction strategy $\theta$, which at step $T$ outputs $\xi_t(\theta)$:
\begin{equation*}
	L_T^{\theta} := \sum_{t=1}^T \lambda (y_t, \xi_t(\theta)) = \sum_{\substack{t = 1, \dots, T:\\{y_t > \xi_t(\theta)}}} \alpha |y_t - \xi_t(\theta)| +\\
	\sum_{\substack{t = 1, \dots, T:\\ y_t < \xi_t(\theta)}} (1-\alpha) |y_t - \xi_t(\theta)| .
\end{equation*}

Our goal is to find a strategy which at time $t$ can compete with any prediction strategy $\xi_t(\theta)$ in terms of cumulative losses.

We denote the {\em regret} at time $T$ to be the difference between the cumulative losses of the learner and the prediction strategy $\theta$:
\begin{equation} \label{eq:regret}
	R_T = L_T - L_T^{\theta},
\end{equation}
and the {\em average regret} at time $T$ to be:
\begin{equation} \label{eq:avg_regret}
	\hat{R}_T = \left(L_T - L_T^{\theta} \right) / T.
\end{equation}

\section{Competitive Quantile Autoregression}
In this section, we describe CQAR, which is built on WAAQR, proposed in \cite{dzhamtyrova2020quantile}, and adapted to time-series forecasting. The algorithm works according to Protocol \ref{protocol1}, which is different from the traditional machine learning approach, where one needs a data set for the algorithm's training. CQAR makes its prediction based on the signal, which is announces by the nature. We assume that the outcomes follow $p$-order autoregressive process (\ref{eq:ar_process}). At the time step $T$ we observe signal $x_T = (1, y_{T-1}, \dots, y_{T-p})$, which contains $p$ previous outcomes. Based on this signal, we need to output the prediction $\gamma_T$ before seeing the true outcome $y_T$. In contrast to QAR, CQAR does not try to find the optimal parameters $\theta$ by minimising the pinball loss function (\ref{eq:min_pinball}).  Instead, CQAR combines the predictions of a large pool of QAR in a way, which is similar to a Bayesian mixture:
\begin{equation}\label{eq:WAAprediction2}
	\gamma_T = \int_{\Theta} \xi_T(\theta) q_{T-1}^{*}(\theta) d\theta,
\end{equation}
where 
\begin{equation}\label{likelihood}
	q_T^*(\theta) = Z q_T (\theta)= Z \exp\Bigl( -\frac{1}{\sqrt{T}} \Bigl(\sum_{\substack{t = 1, \dots, T:\\ y_t < \xi_t(\theta)}}(1-\alpha) |y_t - \xi_t(\theta)| 
	+ \sum_{\substack{t = 1, \dots, T:\\ y_t > \xi_t(\theta)}} \alpha |y_t - \xi_t(\theta)|\Bigr) -a\|\theta\|_1 \Bigr),
\end{equation}
where $a$ is a regularisation parameter and $Z$ is the normalising constant ensuring that $\int_\Theta q_T^{*}(\theta)d\theta = 1$, and $\|\theta\|_1$ denotes $L_1$-norm of parameter $\theta$.
Function $q_T^*(\theta)$ has a meaning of the likelihood of the parameters $\theta$ at time step $T$. 
The pseudo-code of CQAR uses the Metropolis-Hastings algorithm, which is a Markov chain Monte Carlo (MCMC) method \cite{Andrieu2003MCMC}, though any other sampling algorithm could be used instead to approximate the integral (\ref{eq:WAAprediction2}). We start with some initial parameter $\theta^0$ and at each step $m$ we update:
\begin{equation*}
	\theta^m = \theta^{m-1} + \mathcal{N} (0, \sigma^2),~ m=1,\dots, M, 
\end{equation*}
where  $\mathcal{N}(0, \sigma^2)$ is the Gaussian proposal distribution with standard deviation $\sigma$, and $M$ is the total number of MCMC iterations. The Metropolis-Hastings randomly walks through the parameter space $\Theta$, and either accepts or rejects new parameters $\theta$. If the likelihood of the new parameters (\ref{likelihood}) is higher than the old parameters' likelihood, the new parameters are always accepted. Otherwise, the new parameters can be either accepted or rejected. By moving this way, the algorithm mostly samples parameters $\theta$ from the high-density regions of (\ref{likelihood}), only sometimes visiting the area of low-density of the parameters' likelihood. This procedure allows giving an accurate approximation of the integral (\ref{eq:WAAprediction2}).

We provide the pseudo-code of CQAR below. The algorithm has four input parameters: the number of MCMC iterations $M$, the `burn-in period' $M_0$, the regularisation parameter $a$, and the standard deviation $\sigma$. The burn-in period $M_0$ means that we sample $M_0$ values of the parameters, but they are not used in the integral approximation. It is useful as we probably did not yet reach the area of high density of the parameters' likelihood.
\vspace{8pt}
\hrule
\vspace{8pt}
{\bf CQAR}
\vspace{8pt}
\hrule
\vspace{8pt}
{\bf Parameters}: number $M >0$ of MCMC iterations,

\hspace{1.9cm} burn-in period $M_0>0,$

\hspace{1.9cm} standard deviation $\sigma > 0,$

\hspace{1.9cm} regularisation parameter $a > 0$

initialize $\theta_0^M := 0 \in \Theta$ 

define $q_0(\theta) := \exp(-a\|\theta\|_1)$

for $t = 1, 2, \dots$ do

\hspace{0.5cm} $\gamma_t := 0$

\hspace{0.5cm} define $q_{t-1}(\theta) $ by (\ref{likelihood}) if $t>1$

\hspace{0.5cm} read $x_t \in \mathbb{R}^n$

\hspace{0.5cm} initialize $\theta_t^0 = \theta_{t-1}^M$

\hspace{0.5cm} for $m = 1, 2, \dots ,M$ do

\hspace{1cm} $\theta^{*} := \theta_t^{m-1} + \mathcal{N}(0, \sigma^2 I)$

\hspace{1cm} flip coin with success probability

\hspace{1.8cm} $\min \left( 1, q_{t-1}(\theta^{*}) / q_{t-1}(\theta_t^{m-1} ) \right)$

\hspace{1.5cm} if success then

\hspace{2cm} $\theta_t^m := \theta^{*}$

\hspace{1.5cm} else

\hspace{2cm} $\theta_t^m := \theta_{t-1}^m$

\hspace{1.5cm} end if

\hspace{1cm} if $m > M_0$ then

\hspace{1.5cm} $\gamma_t := \gamma_t+ \xi_t(\theta_t^m)$

\hspace{0.5cm} end for 

\hspace{0.5cm} output predictions $\gamma_t= \gamma_t / (M - M_0)$

end for
\vspace{8pt}
\hrule
\vspace{8pt}

An important property of CQAR is that it asymptotically predicts as well as the best QAR. The following theorem provides the upper bound for the average regret between CQAR and the best QAR.
\begin{lemma} {\bf (Theorem 1 in \cite{dzhamtyrova2020quantile})} \label{quantile_bound}
	Let $a > 0$, $A \le y_t \le B$ for any $t=1, 2, \dots, T-1$, where $T$ is a positive integer. For every sequence of outcomes of length $T$, and
	every $\theta \in \mathbb{R}^{p+1}$ 
	the average regret $\hat{R}_T$ between CQAR and QAR satisfies
	\begin{equation*} \label{eq:Loss_quantile}
		\hat{R}_T \le \frac{1}{\sqrt{T}} a \|\theta\|_1 + \frac{1}{\sqrt{T}} \Bigg((p+1) \ln \left(1 + \frac{\sqrt{T}}{a} \max(1, B) \right) + (B-A)^2 \Bigg).
	\end{equation*}
\end{lemma}
The theorem states that CQAR asymptotically predicts as well as the best QAR as the average regret $\hat{R}_T \rightarrow 0$, for $T \rightarrow +\infty$. Although the bound contains the information about the minimum and maximum values of the outcomes at the previous steps, it does not affect the asymptotic behaviour of the bound. The choice of the regularisation parameter $a$ affects the behaviour of the theoretical bound. As a result, it is important to pick the parameter which minimizes the regret's bound. However, in most cases, the optimal choice of the regularisation parameter cannot be found in advance as the number of steps $T$ is usually not known from the start. We discuss the choice of the parameters of CQAR in detail in the experimental part of the article.

\section{Experiments}
The open-source code of our implementation is fully reproducible and available at GitHub~\textsuperscript{\ref{cyber_risk}}. Data is downloaded from the PRC report \textsuperscript{\ref{clearinghouse}}, which contains the chronology of various types of data breaches such as card fraud, insider incidents, paper, and computer physical losses, and unintended information disclosure. The companies which suffer the incidents are classified into seven types of businesses: BSF (Financial and Insurance Services Businesses), BSR (Retail/Merchant including Online Retail Businesses), BSO (Other Businesses), EDU (Educational Institutions), GOV (Government and Military),  MED
(Medical and Healthcare), and NGO (Nonprofits). The report contains 9015 data breaches between January 2005 and September 2019. Analogous to \cite{xu2018breaches}, we focus only on hacking breaches. The total number of observations after removing all incomplete, unknown, and missing breaches are 1602. The data is divided into training and test data sets in the proportion of 60\% to 40\%: the size of the training set is 956, whereas that of the test set is 636. 

\subsection{Data exploration}
We start with data pre-processing. Most days have only one incident per day, 232 days have two incidents, 52 days have three, and 35 days are with more than three incidents. Similarly to  \cite{xu2018breaches}, if several events occur in one day, they are analysed as separate incidents. For these events, we generate a random number from zero to one, which corresponds to some time during the day. After that, these events are sorted by these randomly generated numbers. 

\begin{figure}[ht]
	\centering
	\begin{subfigure}{.5\textwidth}
		\centering
		\includegraphics[width=0.8\linewidth]{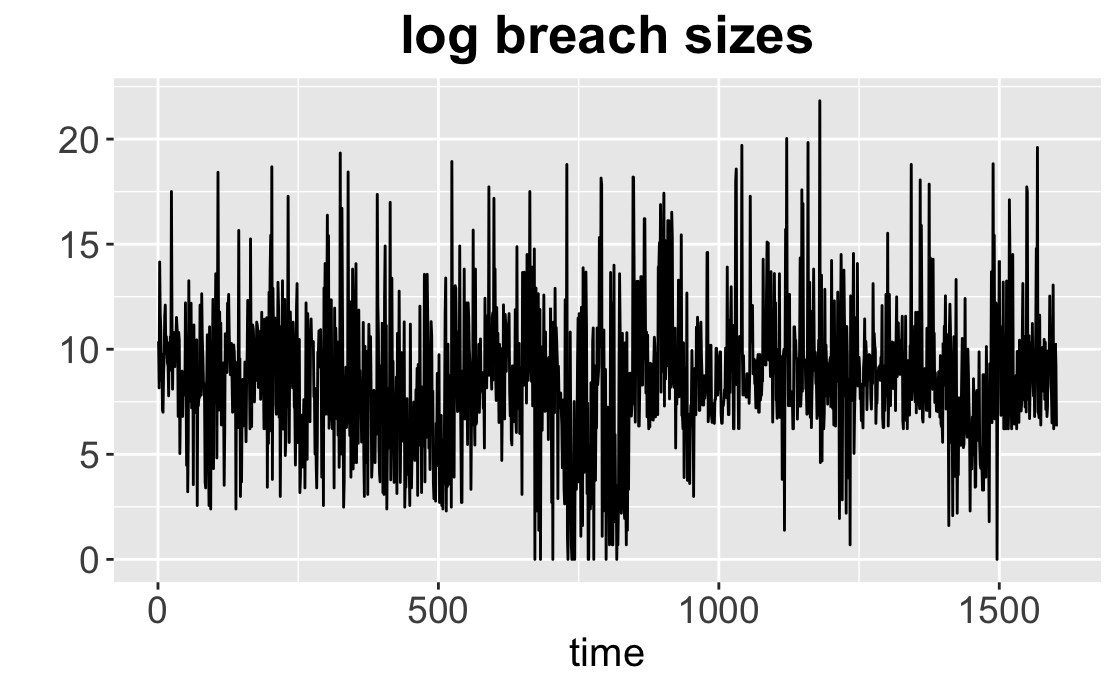}
		\caption{Logarithm of breach sizes}
		\label{fig:vis_size}
	\end{subfigure}%
	\begin{subfigure}{.5\textwidth}
		\centering
		\includegraphics[width=0.8\linewidth]{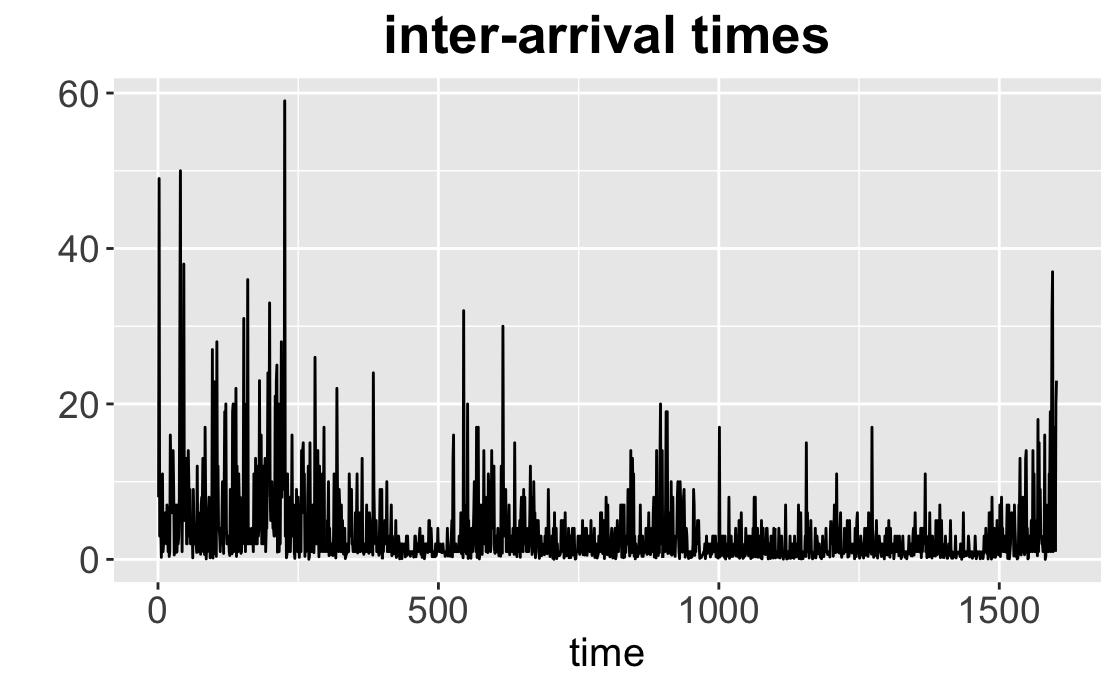}
		\caption{Inter-arrival times}
		\label{fig:vis_times}
	\end{subfigure}
	\caption{Visualisation of breach sizes and inter-arrival times}
	\label{fig:vis}
\end{figure}

Figure \ref{fig:vis} visualises inter-arrival times and the logarithm of breach size, where size is the total number of accounts affected by the breach. We visualise breach sizes on a logarithmic scale because some of the incidents exhibit particularly extreme values. Table \ref{table:stats_sizes} describes the summary statistics of breach sizes, where sd denotes the standard deviation. The analysis in \cite{xu2018breaches} describes the period between January 2005 and April 2017 and contains 600 hacking breaches. We observe that more than 1000 incidents have been added to the report in the last two years. It indicates that either hacking incidents become more frequent or the companies become more transparent about reporting their data breaches. The largest number of incidents are reported in the medical and healthcare sector. The largest incident was reported by Yahoo on the 14th of December 2016, which compromised users' data from three billion accounts.  Table \ref{table:stats_times} shows the same statistics for inter-arrival times. We observe that the mean values of inter-arrival times are less than the standard deviations for each category. It provides evidence that inter-arrival times cannot be modelled with the Poisson distribution. A similar conclusion can be drawn for the breach sizes.

\begin{table}[ht]
	\centering
	\caption{Summary statistics of breach sizes}
	\begin{tabular}{|ccccccc|}
		\hline
		\makecell{type of\\
			organi-\\sation}  & min  & \makecell{median\\ ($\times 10^3$)} & \makecell{mean \\($\times 10^6$)} & \makecell{sd \\($\times 10^6$)} & \makecell{max\\ ($\times 10^6$)} & \makecell{number\\ of obser-\\vations} \\ 
		\hline
		BSF & 6 & 1.7 & 4.8 & 21.3 & 145.5 & 111 \\ 
		BSO & 2 & 10.4 & 26.4 & 214.8 & 3000.0 & 208 \\ 
		BSR & 1 & 2.1 & 6.7 & 33.3 & 327.0& 138 \\ 
		EDU & 12 & 8.5 & 222.5 & 2.7 & 40.0 & 223 \\ 
		GOV & 8 & 6.0 & 457.7 & 2.4 & 21.5 & 93 \\ 
		MED & 1 & 4.0 & 200.1 & 2.9 & 78.8 & 805 \\ 
		NGO & 13 & 4.0 & 142.1 & 0.6 & 3.0 & 24 \\ 
		\hline
		Total & 1 & 4.6 & 4.5 & 78.6 & 3000 & 1602 \\ 
		\hline
	\end{tabular}
	\label{table:stats_sizes}
\end{table}

\begin{table}[ht]
	\centering
	\caption{Summary statistics of breach inter-arrival times}
	\begin{tabular}{|ccccccc|}
		\hline
		\makecell{type of\\
			organi-\\sation}& min & median & mean & sd & max &\makecell{number\\ of obser-\\vations}\\ 
		\hline
		BSF & 0.0111 & 2.00 & 4.16 & 5.78 & 36 & 111 \\ 
		BSO & 0.0480 & 1.00 & 3.08 & 4.18 & 38 & 208 \\ 
		BSR & 0.0233 & 2.00 & 3.52 & 5.09 & 33 & 138 \\ 
		EDU & 0.0134 & 3.00 & 5.86 & 8.12 & 59 & 223 \\ 
		GOV & 0.0842 & 2.00 & 3.66 & 5.06 & 28 & 93 \\ 
		MED & 0.0019 & 1.00 & 2.85 & 4.10 & 37 & 805 \\ 
		NGO & 0.0131 & 1.00 & 2.70 & 3.56 & 13 & 24 \\ 
		\hline
		Total & 0.0019 & 2.00 & 3.49 & 5.20 & 59 & 1602 \\ 
		\hline
	\end{tabular}
	\label{table:stats_times}
\end{table}

\begin{figure}[ht] 
	\centering
	\begin{subfigure}[b]{0.45\linewidth}
		\centering
		\includegraphics[width=0.95\linewidth]{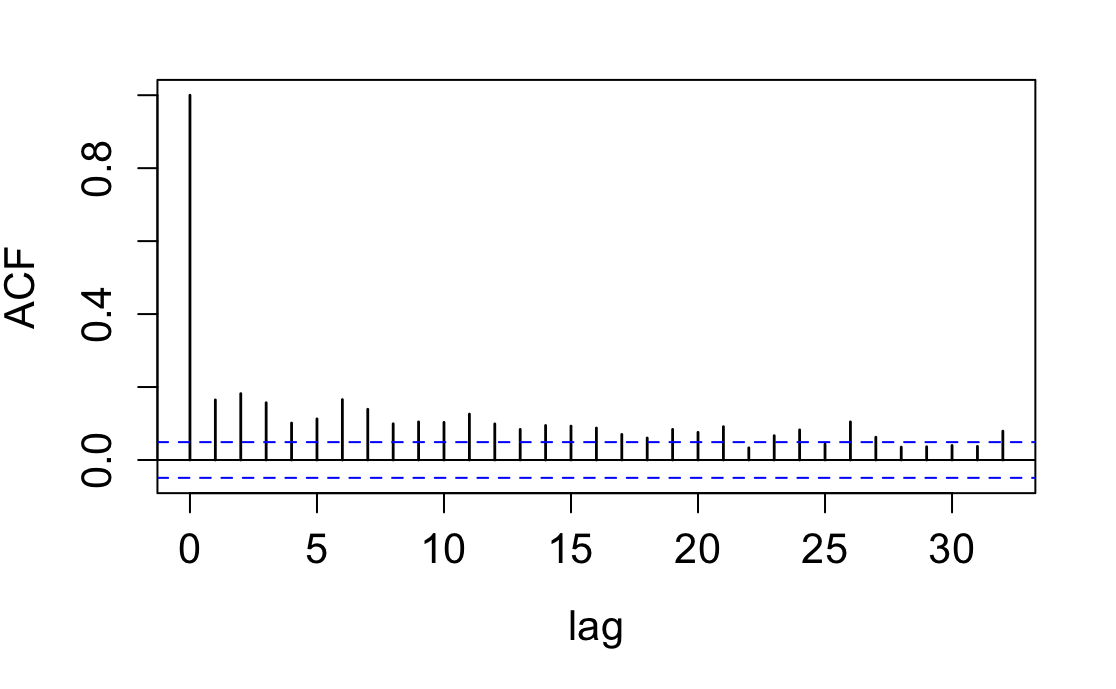} 
		\caption{Logarithm of breach\\ sizes} 
		\label{fig:acf_size} 
		\vspace{4ex}
	\end{subfigure}
	\begin{subfigure}[b]{0.45\linewidth}
		\centering
		\includegraphics[width=0.95\linewidth]{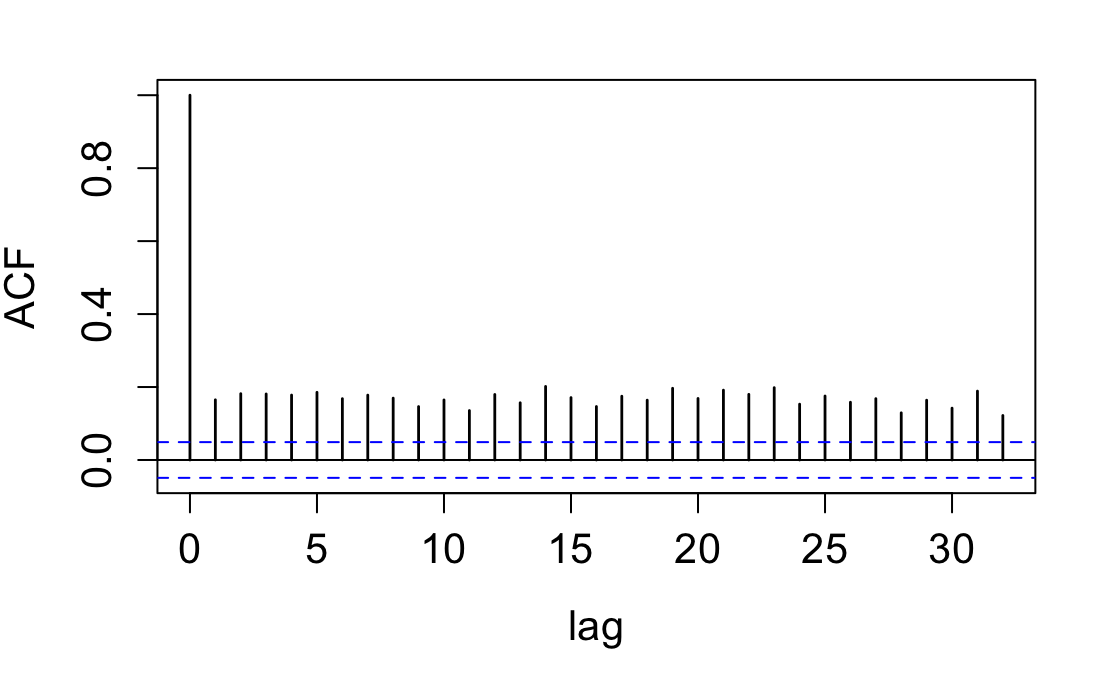} 
		\caption{Logarithm of inter-\\arrival times} 
		\label{fig:acf_time} 
		\vspace{4ex}
	\end{subfigure} 
	\begin{subfigure}[b]{0.45\linewidth}
		\centering
		\includegraphics[width=0.95\linewidth]{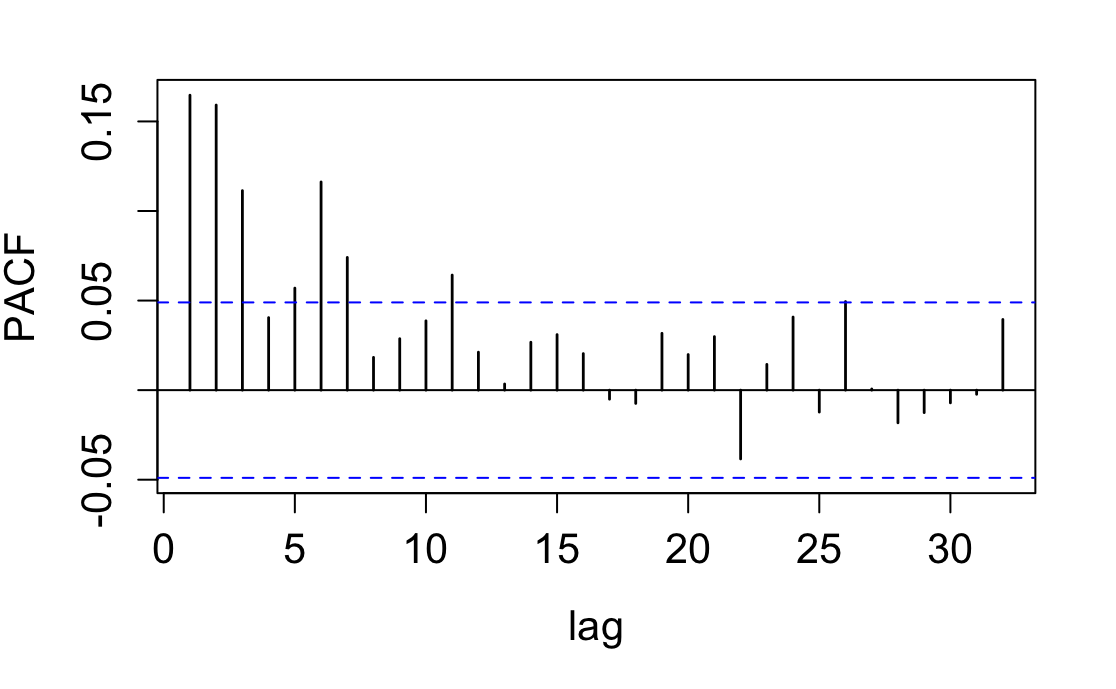} 
		\caption{Logarithm of breach\\ sizes} 
		\label{fig:pacf_size} 
		\vspace{4ex}
	\end{subfigure}
	\begin{subfigure}[b]{0.45\linewidth}
		\centering
		\includegraphics[width=0.95\linewidth]{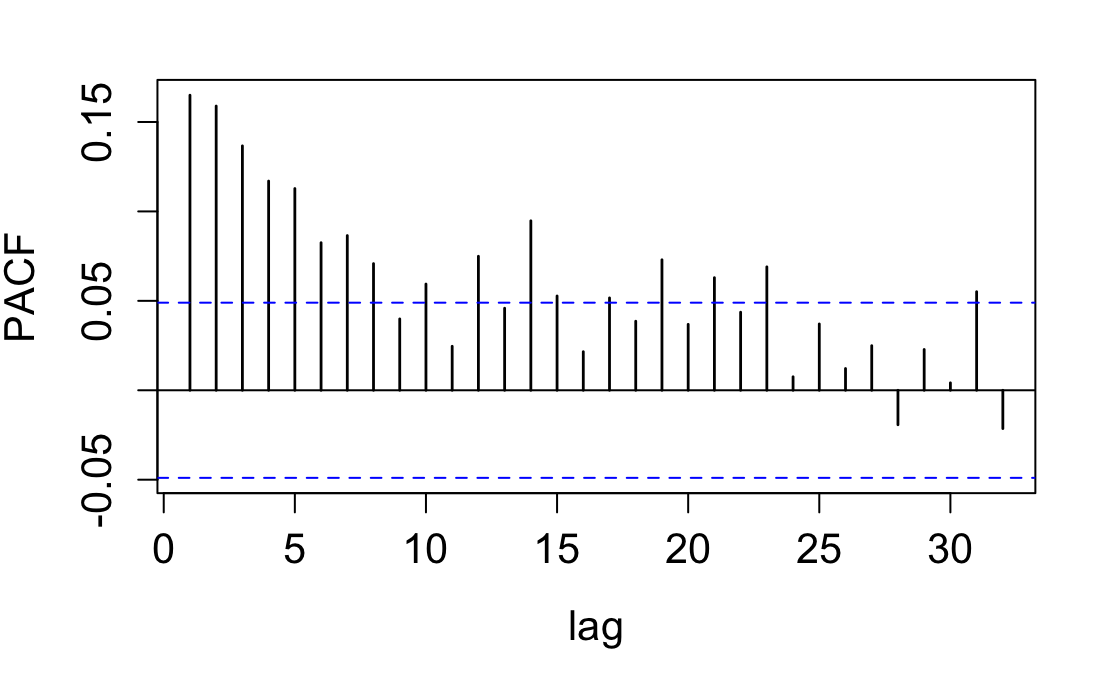} 
		\caption{Logarithm of inter-\\arrival times} 
		\label{fig:pacf_time} 
		\vspace{4ex}
	\end{subfigure} 
	\caption{ACF and PACF}
	\label{fig:acf_pacf}
\end{figure}

Analogously to \cite{xu2018breaches}, we check auto-correlation (ACF) and partial auto-correlation functions (PACF) of the logarithm of breach size and logarithm of inter-arrival time. ACF measures the linear dependence between the lags of time-series, whereas PACF is the correlation between lags adjusted for the contributions of observations in between \cite{hyndman2018forecasting, shumway2016timeseries}. These measures are used to find if observations exhibit a correlation between each other and can be modelled with a stochastic process. Figure \ref{fig:acf_pacf} shows that both breach sizes and inter-arrival times exhibit significant auto-correlations above the threshold values depicted with dotted lines. It indicates that they can be modelled with stochastic processes.

\subsection{Quantile Autoregression}
\begin{figure}[ht]
	\centering
	\begin{subfigure}{.5\textwidth}
		\centering
		\includegraphics[width=0.8\linewidth]{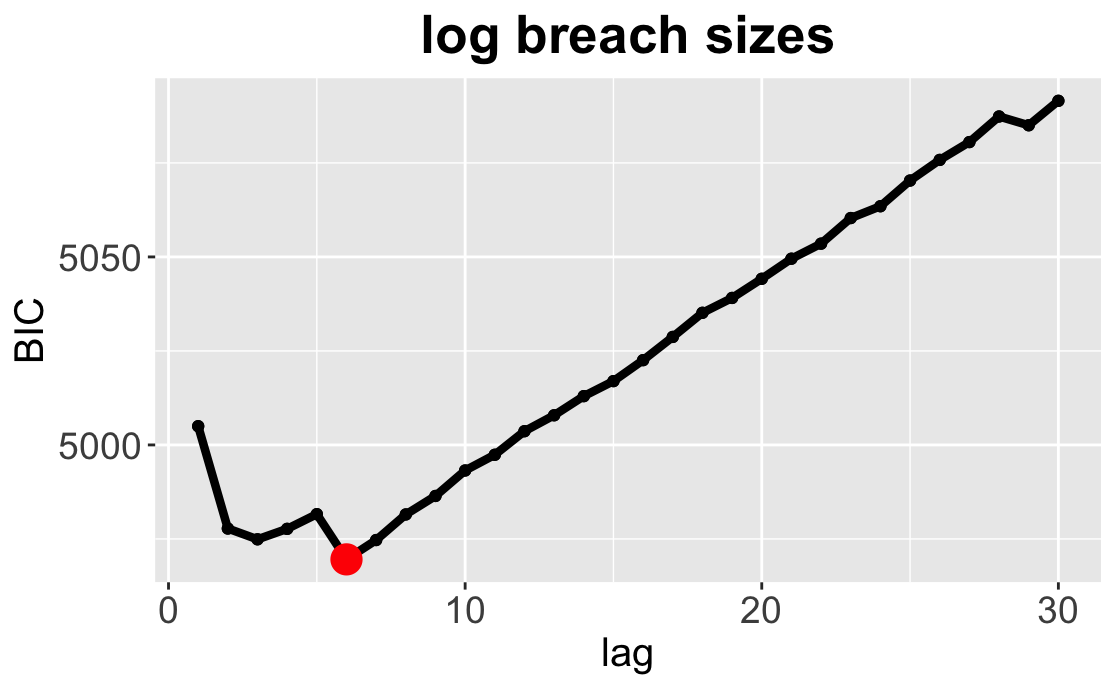}
		\caption{Logarithm of breach sizes}
		\label{fig:bic_size}
	\end{subfigure}%
	\begin{subfigure}{.5\textwidth}
		\centering
		\includegraphics[width=0.8\linewidth]{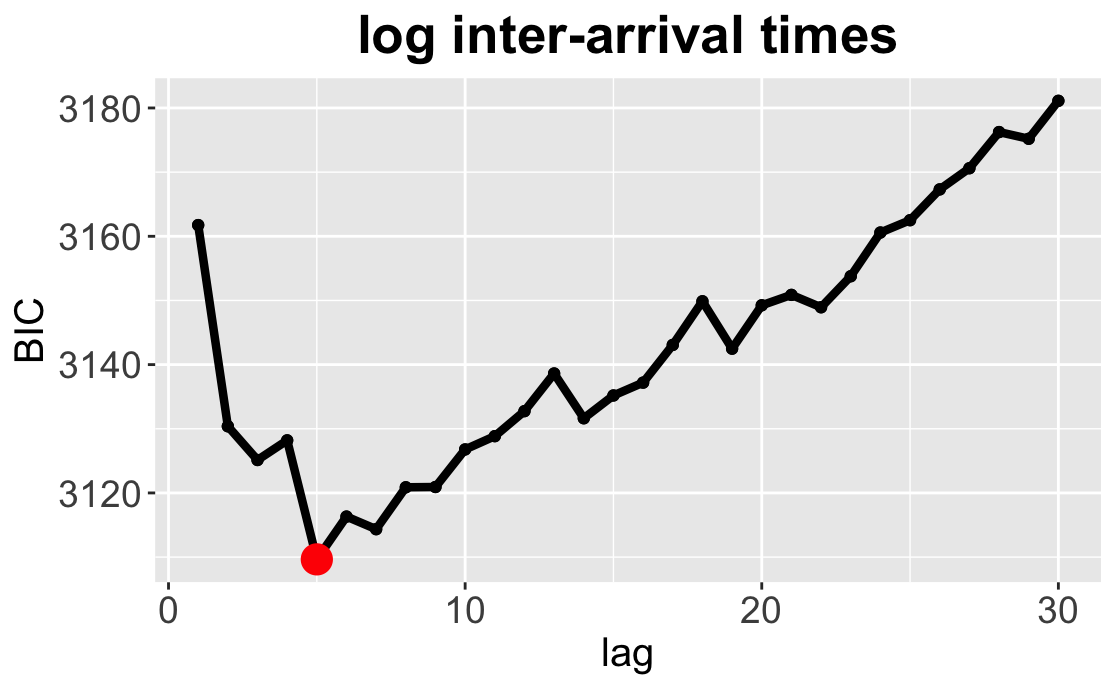}
		\caption{Logarithm of inter-arrival times}
		\label{fig:bic_time}
	\end{subfigure}
	\caption{BIC for different lags}
	\label{fig:bic}
\end{figure}

In this section, we model $\VaR_\alpha$ of the logarithm of breach sizes and the logarithm of inter-arrival times with QAR. First, we need to pick the optimal lag of QAR. Analogous to the problem of choosing the optimal degree of polynomial regression, the optimal order of the autoregressive process (\ref{eq:ar_process}) can be chosen by some information criterion. We use the Bayesian Information Criterion (BIC) \cite{schwarz1978bic} to pick the optimal lag of QAR.
BIC is defined as follows:
\begin{equation*}
	\textrm{BIC} = -2\ln{L} + p \ln N,
\end{equation*}
where $L$ is the maximum of the model's likelihood, $p$ is the number of parameters, and $N$ is the sample size.  BIC penalises complex models with large lag number $p$, and smaller values of the criterion are favourable.
Figure \ref{fig:bic_size} shows BIC values for a different number of lags of QAR, which is built on the training data for quantiles equal to 0.5, i.e.~ median values. The smallest values of BIC correspond to the optimal choice of the lag and are depicted with the red dots. We observe that the optimal values of lag are equal to six in the case of the breach size, and the optimal lag for the inter-arrival time is five. 

\begin{figure}[ht]
	\centering
	\begin{subfigure}{.5\textwidth}
		\centering
		\includegraphics[width=0.8\linewidth]{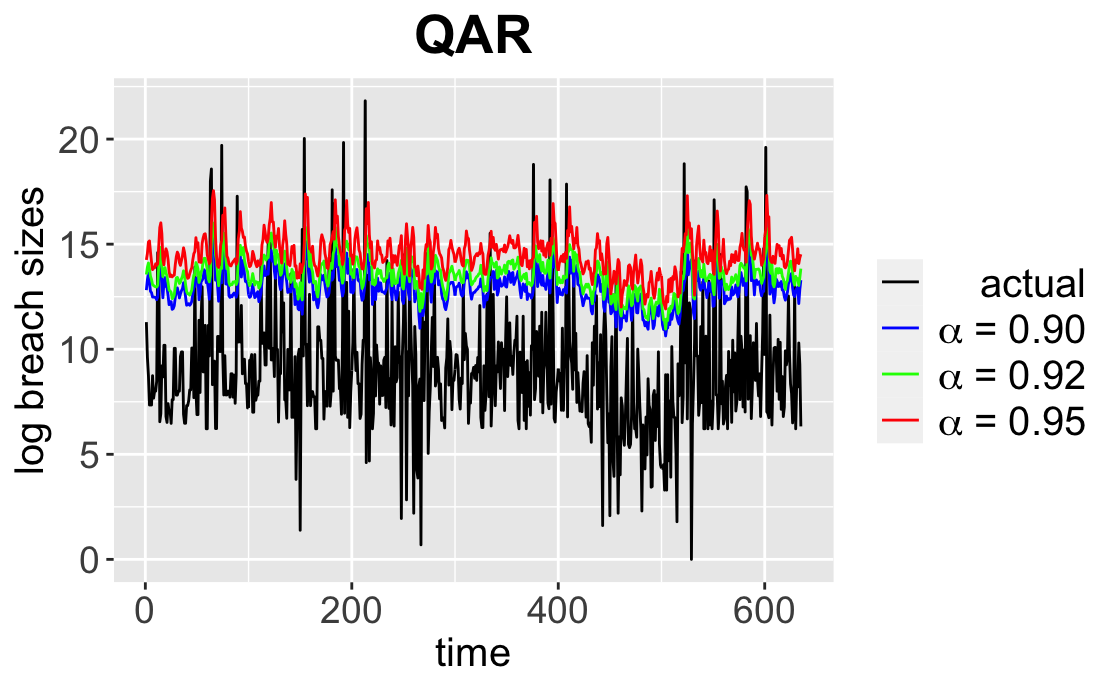}
		\caption{Logarithm of breach sizes}
		\label{fig:qar_size}
	\end{subfigure}%
	\begin{subfigure}{.5\textwidth}
		\centering
		\includegraphics[width=0.8\linewidth]{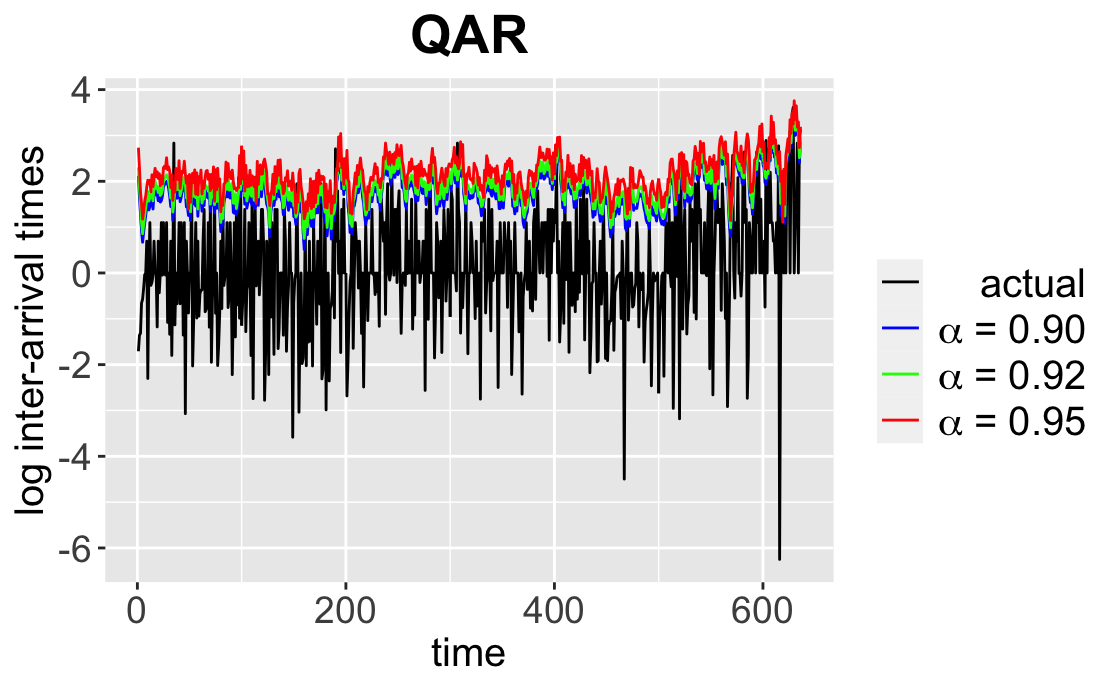}
		\caption{Logarithm of inter-arrival times}
		\label{fig:qar_time}
	\end{subfigure}
	\caption{Predictions of QAR}
	\label{fig:qar}
\end{figure}

We then build QAR for the optimal lags on the training data set. These models are used for making predictions of $\VaR_\alpha$ on the test data set. We pick the significance levels to be $\alpha = 0.9, 0.92, 0.95$. From the risk perspective, it is important to estimate how large the potential losses might be in order to prevent or hedge these losses. Therefore, $\alpha$ values should be large. Figures \ref{fig:qar_size}, \ref{fig:qar_time} illustrate the predictions of QAR for breach sizes and inter-arrival times respectively, on the test data. 

If the observed value exceeds the predicted $\VaR_\alpha$ , we call it {\em violation}. The Kupiec unconditional coverage test \cite{Kupiec1995VaR} measures whether the number of violations is consistent with the confidence level. For example, if $\alpha = 0.9$, then the percent of observation, which exceeds the predicted $\VaR_{0.9}$, should be close to $0.1$. The null hypothesis $H_0$ is that the observed violation rate is equal to $1 - \alpha$. The Kupiec unconditional coverage test focuses only on the number of violations. However, we would like to test whether these exceptions are evenly spread over time. The null hypothesis $H_0$ for the Christoffersen conditional coverage test \cite{Christoffersen1998} is that the probability of observing a violation at some time point does not depend on whether a violation occurred. Table \ref{table:coverage_size} illustrates the results of backtesting of both coverage tests for breach size and inter-arrival time respectively, on the test data. The table shows the expected number of violations of the considered confidence level and the actual number of violations of the considered method. We use the following notations: exp (expected number of violations), act (actual number of violations), the unconditional coverage test p-value (uc.LRp), the conditional coverage test p-value (cc.LRp), the unconditional coverage test decision (uc.D), and the conditional coverage test decision (cc.D), fail to reject the null hypothesis $H_0$ (FR), reject the null hypothesis $H_0$ (R). We can see that QAR(6) fails to reject the null hypothesis $H_0$ for both unconditional and conditional coverage tests, which means that the models fit well and describe the quantiles of breach size correctly. In the case of inter-arrival time, QAR(5) fits well for 0.9 and 0.95 quantiles, however, for 0.92 the conditional coverage test rejects the null hypothesis. In the next section, we show how we can improve the prediction of the breach inter-arrival times by applying CQAR. 
\begin{table}[ht]
	\begin{center}
		\caption{Coverage tests for QAR for breach sizes at test data}
		\begin{tabular}{|cccccccc|}
			\hline
			method &  quantile & exp & act&  uc.LRp & cc.LRp & uc.D & cc.D\\
			\hline
			QAR(6) & 0.90 & 63 & 55 & 0.2509 & 0.4784 & FR & FR \\ 
			QAR(6) & 0.92 & 50 & 44 & 0.3095 & 0.5103 & FR & FR \\ 
			QAR(6) & 0.95 & 31 & 29 & 0.6116 & 0.7456 & FR & FR \\ 
			\hline
		\end{tabular}
		\label{table:coverage_size}
	\end{center}
\end{table}

\begin{table}[ht]
	\begin{center}
		\caption{Coverage tests for QAR for inter-arrival times at test data}
		\begin{tabular}{|cccccccc|}
			\hline
			method &  quantile & exp & act &  uc.LRp & cc.LRp & uc.D & cc.D\\
			\hline
			QAR(5) & 0.90 & 63 & 56 & 0.3062 & 0.3539 & FR & FR \\ 
			QAR(5) & 0.92 & 50 & 41 & 0.1360 & 0.0146 & FR & R \\ 
			QAR(5) & 0.95 & 31 & 26 & 0.2765 & 0.1463 & FR & FR  \\  
			\hline
		\end{tabular}
		\label{table:coverage_time}
	\end{center}
\end{table}

\subsection{Competitive Quantile Autoregression}
In this section, we estimate the hacking breaches' inter-arrival times with CQAR. In contrast to QAR, CQAR does not need a training data set. The algorithm starts its training when it gets the first observation of the test data set. However, as we have the training data set available, we pick the regularisation parameter $a$ and the standard deviation $\sigma$ from the training data. Table \ref{table:params_waaqar} illustrates the acceptance ratio and the total pinball loss of CQAR on the training data set for different parameters  $a$ and $\sigma$. The lowest pinball loss on the training data is achieved with $a = 1$ and $\sigma = 0.7$, which is depicted in bold. The corresponding acceptance ratio for these parameters is 0.27. It is important to `track' the acceptance ratio of CQAR. A very high acceptance ratio might indicate that the algorithm moves too slowly to the optimal parameter $\theta$. Therefore, the total number of iterations and the burn-in period should be increased. Another option is to increase the standard deviation $\sigma$. Table \ref{table:params_waaqar} shows that increasing $\sigma$ leads to decreasing of the acceptance ratio. 
\begin{table}[!htb]
	\caption{Parameters of CQAR on training}
	\begin{subtable}{.45\linewidth}
		\centering
		\caption{Acceptance ratio}
		\begin{tabular}{|c|ccc|}
			\hline
			a \textbackslash  $\hspace{0.1cm} \sigma$ & 0.5 & 0.7 & 1 \\ 
			\hline
			0.1 & 0.69 & 0.47 & 0.22 \\ 
			0.5 & 0.61 & 0.36 & 0.12 \\ 
			1 & 0.53 & 0.27 & 0.06 \\ 
			\hline
		\end{tabular}
	\end{subtable}%
	\begin{subtable}{.45\linewidth}
		\centering
		\caption{Pinball losses}
		\begin{tabular}{|c|ccc|}
			\hline
			a \textbackslash  $\hspace{0.1cm} \sigma$ & 0.5 & 0.7 & 1 \\ 
			\hline
			0.1 & 281.69 & 281.74 & 268.00 \\ 
			0.5 & 177.52 & 171.76 & 172.50 \\ 
			1& 137.20 & \bf{135.42} & 138.21 \\
			\hline
		\end{tabular}
	\end{subtable} 
	\label{table:params_waaqar}
\end{table}

Table \ref{table:coverage_cqar} shows the results of the backtesting for CQAR(5) on the test data set. Note that even though we pick the parameters of the CQAR using the prior knowledge, the algorithm starts with zero parameters $\theta$ and trains using only the test data set. We can see from the table that both unconditional and conditional coverage tests for CQAR(5) fail to reject the null hypothesis. Therefore, CQAR(5) produces better results for predicting breach inter-arrival times than QAR(5). The p-values of CQAR are also higher than p-values of QAR, apart from the cc.LRp for 0.90 quantile.

\begin{table}[ht]
	\begin{center}
		\caption{Coverage tests for CQAR for inter-arrival times at test data}
		\begin{tabular}{|cccccccc|}
			\hline
			method &  quantile & exp & act &  uc.LRp & cc.LRp & uc.D & cc.D\\
			\hline
			CQAR(5) & 0.90 & 63 & 69 & 0.4808 & 0.0785 & FR & FR  \\ 
			CQAR(5) & 0.92 & 50 & 54 & 0.6514 & 0.1025 & FR & FR \\ 
			CQAR(5) & 0.95 & 31 & 27 & 0.3705 & 0.4844 & FR & FR \\ 
			\hline
		\end{tabular}
		\label{table:coverage_cqar}
	\end{center}
\end{table}

\begin{figure}[ht]
	\centering
	\begin{subfigure}{.5\textwidth}
		\centering
		\includegraphics[width=0.8\linewidth]{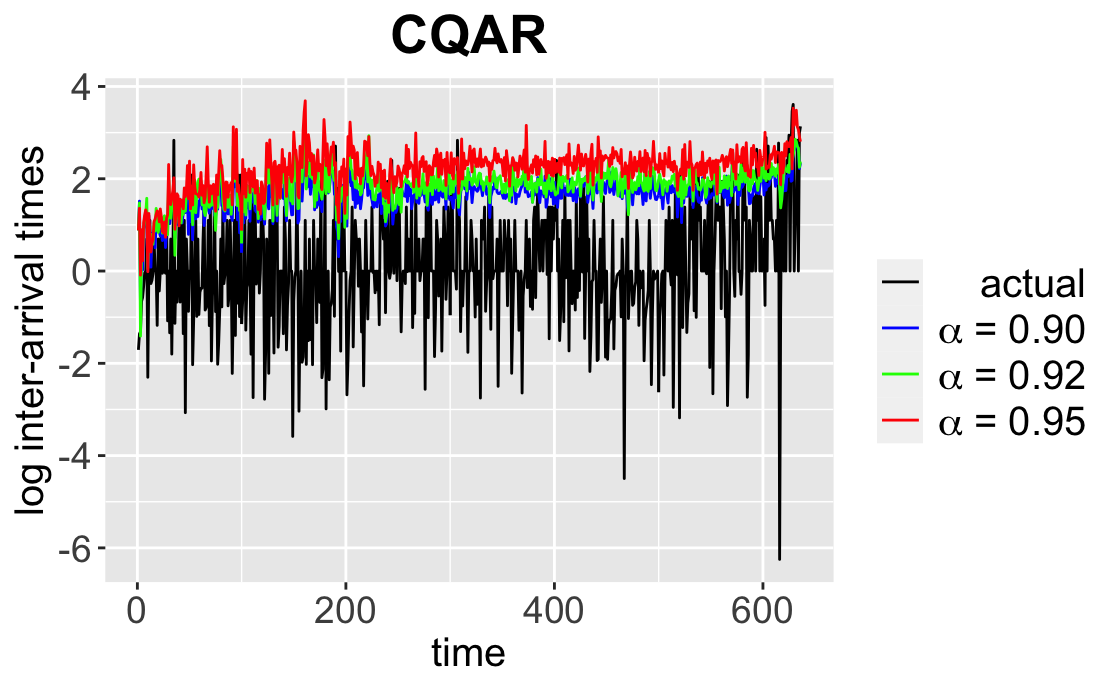}
		\caption{Predictions of CQAR}
		\label{fig:cqar_time}
	\end{subfigure}%
	\begin{subfigure}{.5\textwidth}
		\centering
		\includegraphics[width=0.8\linewidth]{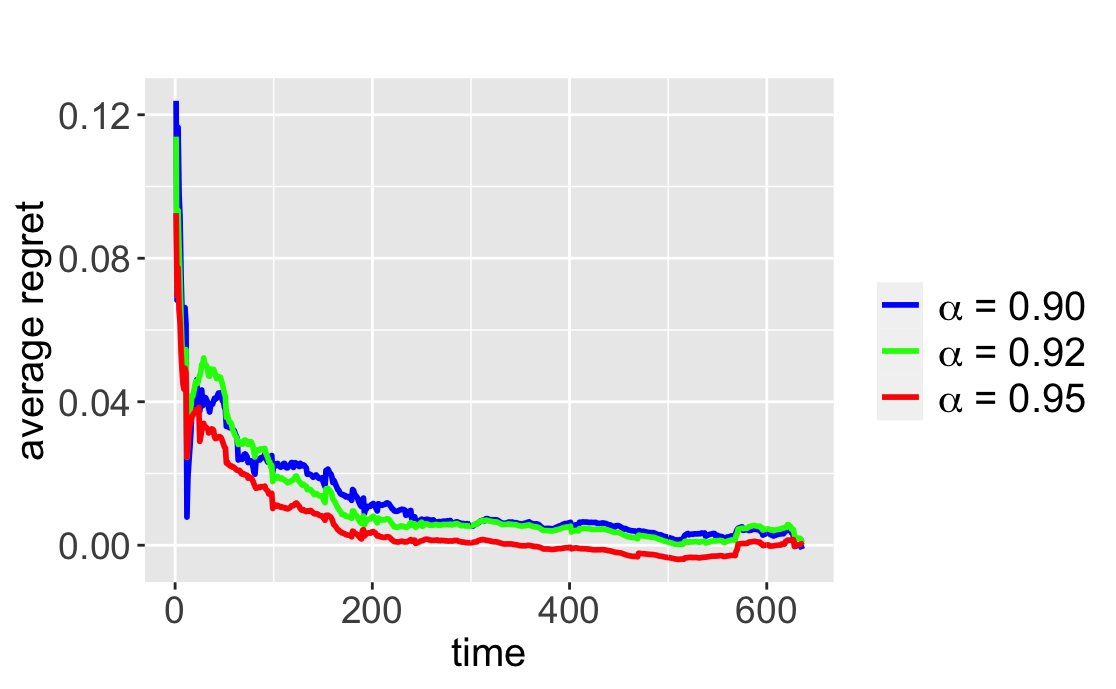}
		\caption{Average regret between CQAR and QAR}
		\label{fig:loss_difference}
	\end{subfigure}
	\caption{CQAR}
	\label{fig:cqar}
\end{figure}

The important property of CQAR is that it asymptotically predicts as well as any QAR.
Figure \ref{fig:cqar_time} illustrates the predictions of CQAR for $\alpha = 0.9, 0.92, 0.95$ on the test data. Figure \ref{fig:loss_difference} shows the average regret between CQAR(5) and QAR(5). As we discussed, CQAR starts with zero parameters $\theta$ at the beginning of its training, and as a result, the average regret is high at the start. However, it becomes close to zero for all considered quantiles as time increases. The resulting graph confirms the theoretical behaviour of the average regret described in Lemma \ref{quantile_bound}.

\section{Conclusions}
In this paper, we have presented two approaches to cyber VaR estimation of time-series. VaR gives a prediction of extreme values with the desired confidence level for a different kind of time-series. These estimates can sequentially be translated into the monetary VaR, which is essential for budget planning and allocation. The first approach to estimate VaR is based on QAR, which provides a new way to model extreme values with the desired confidence level. QAR is more flexible compared to the previously proposed approaches as it allows to model VaR for each confidence level with a separate stochastic process, and hence relies on fewer assumptions on the nature of the data.

The second proposed approach, called CQAR, provides a new framework for dynamic cyber risk estimation. The method re-estimates VaR at each step as soon as new data becomes available. A significant property of this approach is the theoretical guarantee that it asymptotically performs as well as the best QAR found retrospectively. This important property provides confidence in the prediction as it will hold for any new unseen data, while at the same time the method allows adapting to a changing environment. 

Finally, we demonstrate that both methods provide a good fit for predicting the size and inter-arrival times of cyber hacking breaches by running coverage tests. We illustrate the behaviour of the average regret between which conforms to the theoretical bounds of CQAR. In addition, we provide a fully-reproducible code of our experiments.

\section*{Acknowledgments}
This work was supported, in whole or in part, by the Bill \& Melinda Gates Foundation [INV-001309]. Under the grant conditions of the Foundation, a Creative Commons Attribution 4.0 Generic License has already been assigned to the Author Accepted Manuscript version that might arise from this submission.

\bibliographystyle{unsrt}
\bibliography{mylib}

\section*{Appendix}
\begin{lemma} \label{lemma:WAAloss} {\bf(Lemma 2 in \cite{Levina2010WAA})}
	Let $\lambda(y, \gamma) \le L$ for all $y \in \Omega$ and $\gamma \in \Gamma$. The Weak Aggregating Algorithm guarantees that, for all $T$ 
	\begin{equation*}
		L_T \le \sqrt{T} \left( -\ln \int_{\Theta} \exp \left(-\frac{L_T^{\theta}}{\sqrt{T}} \right) P_0 (d\theta) + L^2 \right).
	\end{equation*}
\end{lemma}
{\bf Proof of Lemma \ref{quantile_bound}}
\begin{proof}
	The proof is the same as the proof of Theorem 1 from \cite{dzhamtyrova2020quantile} adjusted for the time-series data. We provide the proof here for completeness. 
	
	Recall, that the learner makes a prediction $\gamma_t$ based on the signal $x_t = (1, y_{t-1}, \dots, y_{t-p}) \in \mathbb{R}^{p+1}$, and outcomes come from the interval $[A, B]$. We choose an initial distribution of parameters
	\begin{equation}\label{eq:weights_init}
		P_0(d\theta) = \left(\frac{a}{2} \right)^{p+1} e^{-a \|\theta\|_1}d\theta,
	\end{equation}
	for some $a>0$, and $\theta \in \mathbb{R}^{p+1}$.
	Let us define the truncated expert $\tilde{\mathcal{E}}_{\theta}$ which at step $t$ outputs:
	\begin{equation} \label{experts_trunc}
		\tilde{\xi}_t(\theta) = \begin{cases}
			A, & \text{if $x_t^\prime \theta < A$}  \\
			x_t^\prime \theta, & \text{if $A \le x_t^\prime \theta \le B$}  \\
			B, & \text{if $x_t^\prime \theta > B$} 
		\end{cases}.
	\end{equation}
	Let us denote $\tilde{L}^{\theta}_T$ the cumulative loss of expert $\tilde{\mathcal{E}}_{\theta}$ at the step $T$:
	\begin{equation}\label{loss_experts_trunc}
		\tilde{L}^{\theta}_T := \sum_{t=1}^T \lambda(y_t, \tilde{\xi}_t (\theta)).
	\end{equation}
	We apply the algorithm's prediction (\ref{eq:WAAprediction2}) with truncated experts $\tilde{\mathcal{E}}_{\theta}$. As experts output predictions inside the interval $[A, B]$, and the algorithm's prediction is a weighted average of experts' predictions, then $\gamma_t$ lies in the interval $[A, B]$, $\forall t$.
	
	We can bound the maximum loss at each time step:
	\begin{equation} \label{eq:Lmax}
		L:= \max_{y \in [A,B],~\gamma \in [A, B]} \lambda(y, \gamma) \le (B-A) \max(\alpha, 1-\alpha) \le B-A.
	\end{equation}
	
	Lemma \ref{lemma:WAAloss} provides the theoretical guarantees for the strategy that follows WAA (\ref{eq:WAAprediction2}) used by our algorithm. Applying Lemma \ref{lemma:WAAloss} for initial distribution (\ref{eq:weights_init}) and putting the maximum loss (\ref{eq:Lmax}) we obtain:
	\begin{equation} \label{eq:LossWAA}
		L_T \le \sqrt{T}  \left( -\ln \left( \left( \frac{a}{2} \right)^{p+1}  \int_{\mathbb{R}^{p+1}} e^{- \tilde{J}(\theta)} d\theta\right)+ (B-A)^2 \right),
	\end{equation}
	where
	\begin{equation} \label{eq:J}
		\tilde{J}(\theta) := \frac{\tilde{L}_T^{\theta}}{\sqrt{T}} + a \|\theta\|_1.
	\end{equation}
	
	For all $\theta, \theta_0 \in \mathbb{R}^{p+1}$ we have:
	\begin{multline}\label{loss_part1}
		\sum_{\substack{t = 1, \dots, T:\\ y_t < x_t^\prime \theta}} |x_t^\prime \theta - y_t| \le \sum_{\substack{t = 1, \dots, T:\\ y_t < x_t^\prime \theta}}  |x_t^\prime \theta_0 - y_t| 
		+ \sum_{\substack{t = 1, \dots, T:\\ y_t < x_t^\prime \theta}}  |x_t^\prime \theta - x_t^\prime \theta_0| \\
		\le \sum_{\substack{t = 1, \dots, T:\\ y_t < x_t^\prime \theta}}  |x_t^\prime \theta_0 - y_t| + \sum_{\substack{t = 1, \dots, T:\\ y_t < x_t^\prime \theta}}  \smash{\displaystyle\max_{t=1,\dots,T}} \|x_t\|_{\infty} \|\theta - \theta_0\|_1 \\
		\le \sum_{\substack{t = 1, \dots, T:\\ y_t < x_t^\prime \theta}}  |x_t^\prime \theta_0 - y_t| + T \max(1, B) \|\theta - \theta_0\|_1.
	\end{multline}
	
	Analogously, we have:
	\begin{equation}\label{loss_part2}
		\sum_{\substack{t = 1, \dots, T:\\ y_t > x_t^\prime \theta}}  |x_t^\prime \theta - y_t| 
		\le \sum_{\substack{t = 1, \dots, T:\\ y_t > x_t^\prime \theta}}  |x_t^\prime \theta_0 - y_t| \\+ T \max(1, B)  \|\theta - \theta_0\|_1.
	\end{equation}
	
	By multiplying inequality (\ref{loss_part1}) by $(1-\alpha)$, inequality (\ref{loss_part2}) by $\alpha$ and summing them, we have:
	\begin{equation} \label{loss_ineq}
		L_T^{\theta} \le L_T^{\theta_0} + T  \max(1, B) \|\theta - \theta_0\|_1.
	\end{equation}
	
	The cumulative loss of truncated expert $\tilde{\mathcal{E}}_{\theta}$  cannot exceed the cumulative loss of non-truncated expert $\mathcal{E}_{\theta}$ for all $\theta \in \mathbb{R}^{p+1}$:
	\begin{equation} \label{loss_ineq2}
		\tilde{L}_T^{\theta} \le L_T^{\theta} .
	\end{equation}
	
	By dividing (\ref{loss_ineq}) by $\sqrt{T}$ and adding $a \|\theta\|_1$ to both parts, we have:
	\begin{multline}
		\tilde{J}_T(\theta) \le J_T(\theta) \le J_T(\theta_0) + \sqrt{T}  \max(1, B) \|\theta - \theta_0\|_1  +  a \left(\|\theta\|_1 - \|\theta_0\|_1 \right) \\
		\le J_T(\theta_0) + \left(\sqrt{T}  \max(1, B) + a\right) \|\theta - \theta_0\|_1,
	\end{multline}
	where 
	\begin{equation} \label{eq:J2}
		J_T(\theta) := \frac{L_T^{\theta}}{\sqrt{T}} + a \|\theta\|_1.
	\end{equation}
	
	Let us denote $b_T =\sqrt{T} \max(1, B) + a$. We evaluate the integral:
	\begin{multline*}\label{integral}
		\int_{\mathbb{R}^{p+1}} e^{- \tilde{J}_T(\theta)} d\theta \ge \int_{\mathbb{R}^{p+1}} e^{ -  (J_T(\theta_0) + b_T \|\theta - \theta_0\|_1 )} d\theta 
		= e^{-J_T(\theta_0)} \int_{\mathbb{R}} \hdots \int_{\mathbb{R}} e^{- b_T \sum_{i=1}^{p+1}|\theta_i - \theta_{i,0}| } d\theta_i  \\
		= e^{-J_T(\theta_0)} \int_{\mathbb{R}} \hdots \int_{\mathbb{R}}  \prod_{i=1}^{p+1} e^{- b_T |\theta_i - \theta_{i,0}|} d\theta_i 
		= e^{-J_T(\theta_0)} \prod_{i=1}^{p+1} \int_{\mathbb{R}} e^{- b_T |\theta_i - \theta_{i,0}|} d\theta_i
		= e^{- J_T(\theta_0)} \left( \frac{2}{ b_T}\right)^{p+1}.
	\end{multline*}
	By putting this expression in (\ref{eq:LossWAA}) we obtain:
	\begin{equation*} 
		L_T \le L_T^{\theta} +\sqrt{T} a \|\theta\|_1 + \sqrt{T} \Bigg((p+1) \ln \left(1 + \frac{\sqrt{T}}{a} \max(1, B) \right) + (B-A)^2 \Bigg).
	\end{equation*}
	By putting this expression in formula for the average regret (\ref{eq:avg_regret}) we obtain the theoretical bound from Lemma \ref{quantile_bound}.
\end{proof}
\end{document}